\documentstyle[eqsecnum,aps,epsfig]{revtex}
\begin{document}
\preprint{hep-ph/0104165}
\draft
\title{
Hint for axial-vector contact interactions in the data on\\
${\bf e^+e^-\rightarrow e^+e^-(\gamma)}$ at centre-of-mass energies
192--208 GeV
}
\author{D. Bourilkov}
\address{
Institute for Nuclear Research and Nuclear Energy,
blvd. Tzarigradsko chaussee 72, BG-1784 Sofia, Bulgaria\\
and Institute for Particle Physics (IPP), ETH Z\"urich,
CH-8093 Z\"urich, Switzerland
}
\date{\today}
\maketitle
\begin{abstract}
For the first time the experiments ALEPH, DELPHI, L3 and OPAL
have presented preliminary results for fermion-pair
production in $e^+e^-$ collisions on the full data set above the Z pole.
A combined analysis of the Bhabha scattering
measurements is performed to search for effects of
contact interactions. In the case of two axial-vector (AA)
currents the best fit to the data is 2.6 standard deviations
away from the Standard Model expectation, corresponding
to an energy scale $\Lambda = 10.3^{+2.8}_{-1.6}$~TeV for contact
interactions.
For other models no statistically significant deviations are observed,
and the data are used to set lower limits at 95~\% confidence level
on the contact interaction scales ranging from 8.2 to 21.3 TeV,
depending on the helicity structure.
\end{abstract}


\section{Introduction}
\label{sec:intro}

The Standard Model (SM) is very successful in confronting the
data coming from the highest energy accelerators.
Still, there are theoretical reasons to expect that it is 
an effective theory, valid in a limited energy range, and
one of the first questions in the quest for physics beyond the
Standard Model is what is the relevant scale, where new phenomena
will give experimental signatures. In this paper we will follow
a data-driven approach and analyze the full set of measurements
on the reaction
\begin{equation}
e^+e^-\rightarrow e^+e^-(\gamma)
\label{eq0}
\end{equation}
at centre-of-mass energies $\sqrt{s}$ above the Z resonance
from 183 up to 208 GeV. Rather than concentrating on a
particular model, we will look for something unexpected. 
Bhabha scattering is chosen as a very sensitive probe, affected in many
new physics scenarios. This analysis is a continuation of the work
presented in~\cite{Bourilkov:2000}, based on the published differential cross
sections for Bhabha scattering at energies 183 and 189 GeV.

In the Standard Model the production of a fermion-pair $f\bar f$
in $e^+e^-$ collisions
is described by the exchange of $\gamma$ or Z in the $s$-channel, and if the
final state is identical to the initial one, also in the $t$-channel.
The interest in studying fermion-pair final states above the Z pole
at LEP2 is driven by the fact that many types of new physics scenarios can contribute
to these processes. For this to 
happen, the couplings to the initial and final states should be different
from zero. In the case of Bhabha scattering we just need a coupling to
the electron and the positron. Even if the Standard Model extension operates
at an energy scale much higher than the accessible centre-of-mass energy
for a direct observation, it can still give measurable effects by modifying
the differential cross section through interference with the SM amplitudes.

\section{Contact Interactions}
\label{sec:ci}

Contact interactions offer a general framework for describing 
a new interaction with typical energy scale  $\Lambda \gg \sqrt{s}$.
The presence of operators with canonical dimension
$N > 4$ in the Lagrangian gives rise to effects  $\sim 1/\Lambda^{N-4}$.
Such interactions can occur for instance, if the SM particles are composite, or
when new heavy particles are exchanged.

For fermion-pair production, the lowest order flavor-diagonal and
helicity-conserving operators have dimension six~\cite{PeskinCI}.
The differential cross section takes the form
\begin{equation}
\frac{d \sigma}{d \Omega} = SM(s,t)+\varepsilon\cdot C_{Int}(s,t)+\varepsilon^2\cdot C_{CI}(s,t)
\end{equation}
where the first term is the Standard Model contribution, the second comes from
interference between the SM and the contact interaction, and the third is the
pure contact interaction effect.
The Mandelstam variables are denoted as $s$, $t$ and $u$.
Usually the coupling is fixed, $\frac{g^2}{4\pi}=1$,
and the structure of the interaction is parametrized by coefficients
for the helicity amplitudes:
$|\eta_{ij}|\leq 1$, where $(i,j=\mathrm{L,R})$ labels
the helicity of the incoming and outgoing fermions.
We define 
\begin{equation}
\varepsilon = \frac{g^2}{4\pi}\frac{{\rm sgn}(\eta)}{\Lambda^2}
\end{equation}
where the sign of $\eta$ enables to study both the cases of positive
and negative interference.

\section{Experimental Data}
\label{sec:edata}

Measurements on fermion-pair production for the full
data set of the LEP2 collider have become available recently.
They include preliminary data from 192 to 208 GeV centre-of-mass energies
and published results at 183 and 189 GeV.
In the following we will concentrate on the measurements of
Bhabha scattering, where large data samples have been
accumulated during the very successful LEP runs from 1997 to 2000.

The ALEPH~\cite{al183,al189,al202,al209},
DELPHI~\cite{de189,de202,de209},
L3~\cite{l3189,l3202,l3209}
and OPAL~\cite{op183,op189,op202,op209}
collaborations have presented results for the total cross
section and the forward-backward asymmetry $A_{FB}$ of Bhabha scattering.
In the case of DELPHI, L3 and OPAL the results are for all
energy points and the scattering angle $\theta$ is
the angle between the incoming and the outgoing electrons
in the laboratory frame.
In the ALEPH case the forward-backward asymmetry is available only at 183 GeV
and the scattering angle $\theta^{*}$ is defined
in the outgoing $e^+e^-$ rest frame.
The acceptance is given by the angular range
$-0.9 < \cos\theta^{*} < 0.7$ for ALEPH, by
$44^{\circ}<\theta<136^{\circ}$ for DELPHI and L3, and by
$|\cos\theta|<0.7$ for OPAL.

The experiments use different strategies to isolate the
high energy sample, where the interactions take place at energies
close to the full available centre-of-mass energy. This
sample is the main search field for new physics~\footnote{
For reviews see e.g.~\cite{Bourilkov:1998,Bourilkov:2001}.}.
DELPHI, L3 and OPAL apply an acollinearity cut of $20^{\circ}$, $25^{\circ}$
and $10^{\circ}$ respectively.
ALEPH defines the effective energy, $\sqrt{s'}$, as the invariant
mass of the outgoing fermion pair. It is determined from the
angles of the outgoing fermions.
For details of the selection procedures, the statistical and
systematic errors we refer the reader to the publications of
the LEP experiments.

\section{Analysis Method}
\label{sec:method}

The Standard Model predictions for Bhabha scattering are computed with
the Monte Carlo generator {\tt BHWIDE}~\cite{BHWIDE}.
We assign a theory uncertainty of 1.5~\% on the absolute scale
of the predictions~\cite{LEP2MC}.
In all cases the individual experimental
cuts of the selection procedures and the isolation of the high
energy samples are taken into account.
The results are cross-checked with the semi-analytic program
{\tt TOPAZ0}~\cite{TOPAZ0}
and the Monte Carlo generator {\tt LABSMC}~\cite{LABSMC}.

The effects of new phenomena are computed as a function of
the parameter $\varepsilon$, defined in Section~\ref{sec:ci}.
Initial-state radiation (ISR) changes the effective centre-of-mass
energy in a large fraction of the observed events.
We take these effects into account by computing the first order
exponentiated cross section following~\cite{Kleiss}.
Other QED and electroweak corrections give smaller effects and are neglected.

In total we have 57 data points: 32 from the cross sections
(eight energy points and four experiments) and 25 from the
forward-backward asymmetries.
A fitting procedure similar to the one
in~\cite{Bourilkov:1999,sneut} is applied.
A negative log-likelihood function is constructed by combining
all data points at the eight centre-of-mass energies:
\begin{equation}
-\log {\cal{L}} = \rm \sum_{r=1}^{n}\left(\frac{[prediction(SM, \varepsilon) - measurement]^2}{2 \cdot \Delta^2}\right)_r
\label{eqll1}
\end{equation}
where $\rm prediction(SM, \varepsilon)$ is the SM expectation for a
given measurement (total cross section or forward-backward asymmetry)
combined with the additional effect of contact interactions as a function of
the scale $\Lambda$,
$\rm measurement$ is the corresponding measured quantity and
\mbox{$\Delta = \rm error[prediction(SM, \varepsilon) - measurement]$}.
The index $r$ runs over all data points.
The error on a deviation consists of three parts, which are combined in
quadrature: a statistical error and a systematic error (as given by the
experiments) and the theoretical error assigned above.
The systematic errors account for small correlations between
data points. The minimum of the negative log-likelihood function
gives the central value of the fitted parameter $\varepsilon$
for each model, and the interval containing 68\% of the total
probability around the minimum is used to determine the values corresponding
to one standard deviation in the positive and negative directions.

\section{Results and Discussion}
\label{sec:results}

\begin{figure}
  \begin{center}
\mbox {\epsfig{file=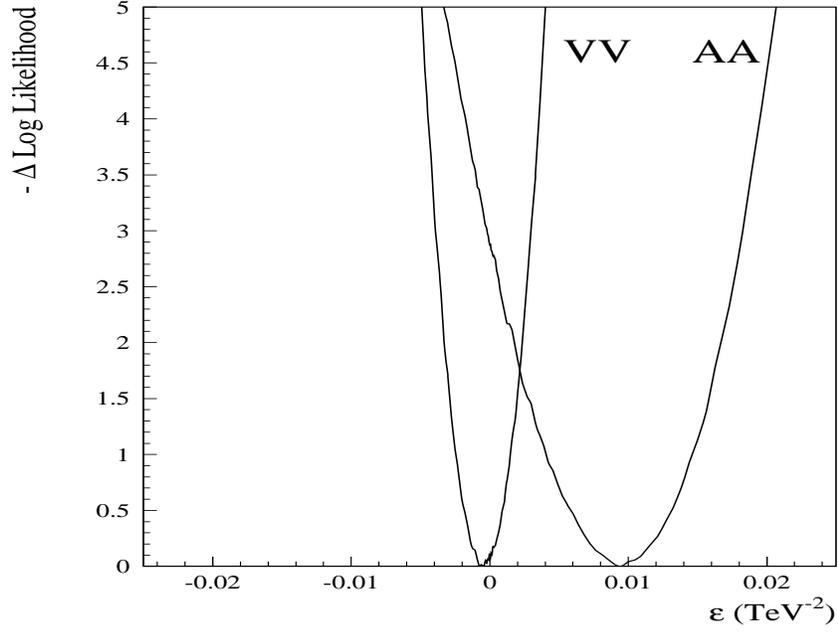,width=0.70\textwidth,height=0.40\textheight}}
  \end{center}
\caption{Log-likelihood curves for the VV and AA contact interaction models
         of the combined fits to the data on Bhabha scattering from the
         four LEP experiments.}
\label{figure1}
\end{figure}

The results of the fits for the different contact interaction models
are summarized in Table~\ref{tab:ci-leptons}. In the same table the
coefficients specifying the helicity structure of the investigated
models are given.
For all models the central value of the parameter $\varepsilon$ is
no more than 1.05 standard deviations away from the Standard Model
value $\varepsilon = 0$, with one exception. The fitted value for the
AA model is
\begin{equation}
\varepsilon = 0.0095\ ^{+ 0.0036}_{- 0.0037} \ \ \rm TeV^{-2}
\label{eqaa}
\end{equation}
or 2.6 standard deviations from the SM expectation. The quality of all
fits is good, with $\chi^2$ values $\sim 52/56$~degrees of freedom.
For the AA model the $\chi^2$ value is $46.8/56$~degrees of freedom.
The log-likelihood curves for the VV and AA models
are shown in Fig.~\ref{figure1}.

In the cases where the data from the LEP collaborations 
show no statistically significant deviations from the SM
predictions, we can derive one-sided lower limits on the scale
$\Lambda$ of contact interactions at 95\% confidence level.
This is done by integrating the log-likelihood functions
in the physically allowed range of the parameters describing
new physics phenomena, assuming a uniform prior distribution.
The exact definition can be found in~\cite{Bourilkov:1999}.
The limits for positive or negative interference are
summarized in Table~\ref{tab:ci-leptons}.
The results presented here improve on the limits obtained by
individual LEP experiments~\cite{al183,de189,l3ci,l3189newph,op189}
or in a combined analysis~\cite{Bourilkov:2000}.

\begin{figure}
  \begin{center}
\mbox {\epsfig{file=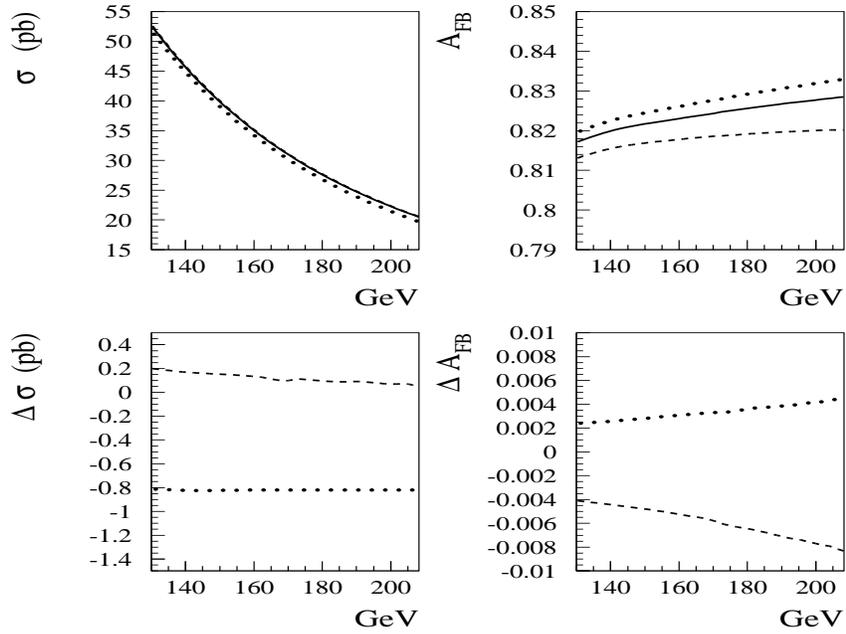,width=0.70\textwidth,height=0.40\textheight}}
  \end{center}
\caption{Effects of contact interactions on the total cross section and
         the forward-backward asymmetry for Bhabha scattering as a function
         of the centre-of-mass energy (upper plots) and difference to the
         Standard Model predictions (lower plots): SM - solid, 
         VV model - dotted and AA model - dashed. The energy scale is
         $\Lambda_+ = 10.3$~TeV in both cases. The angular range is
         $44^{\circ}<\theta<136^{\circ}$.}
\label{figure2}
\end{figure}

Now we turn to a discussion of the result for the AA model.
Clearly this result is unexpected and requires careful
analysis. The central value and the one standard deviation
band of Eq.~\ref{eqaa} correspond to a scale for axial-vector
contact interactions
\begin{equation}
\Lambda_+ = 10.3 \ \left\{^{13.1 \ \ (-\,1\ \sigma)}_{\ 8.7\ \ (+\,1\ \sigma)}\right. \ \ \rm TeV.
\label{eqlambdaaa}
\end{equation}
Let us use this central value in our investigation. We can reverse
the view point and ask ourselves what are the expected deviations
from the SM predictions for the total cross section and the
forward-backward asymmetry of Bhabha scattering. The result
as a function of the centre-of-mass energy is illustrated
in Fig.~\ref{figure2}. The AA model changes the total cross
section by less than 0.5\% in this energy range. The relative effect is
practically constant for rising energy. For comparison, the combined
error on the total cross section of the four experiments for the 
highest energy point (207 GeV) is 1.2\%.
An additional factor is the theoretical uncertainty of 1.5\% on the
absolute scale, discussed in Section~\ref{sec:method}.
On the contrary, the forward-backward asymmetry is changed by (-0.0066) in
absolute value (or -0.8\% relative) at 183 GeV and by (-0.0083) or (-1.0\%) at
207 GeV. The effect is rising with energy. It is clear that the
AA model manifests itself mainly in the forward-backward asymmetry,
which is not affected by the uncertainty on the absolute scale.
Another favorable fact is that the statistical error on $A_{FB}$
is given by $\sigma(A_{FB}) = \sqrt{(1-A_{FB}^2)/N}$, so for the
same number of events $N$ we have a reduction factor ($\sim$~0.6) due to the
large value of the asymmetry $\sim$~0.8. On the minus side this
measurement requires recognition of the electron and positron
charges, and hence stricter event selection asking for good tracks.
Also corrections for charge confusion have to be applied.
For comparison, the combined error on the forward-backward asymmetry
for three experiments and the highest energy point (207 GeV) is 1.1\%.
As the ALEPH collaboration has not presented results on $A_{FB}$ above
183 GeV, the result discussed here is based on the asymmetry measurements
of the other three collaborations.

The measurements of Bhabha scattering from 192 to 208 GeV are
{\em preliminary}. The systematic errors and the central values
may still change. The charge confusion correction has the form
\begin{equation}
A_{FB}^{{\rm corrected}} = \frac{A_{FB}^{{\rm measured}}}{1-2c}
\end{equation}
where c is the amount of events with wrong sign assignment.
For example, if $c = 0.05$, it is enough to underestimate
the charge confusion by 9\% in order to underestimate the
asymmetry by 1\%. 
For $c = 0.02$, one has to underestimate the charge confusion 
already by 24\% in order to underestimate the asymmetry by 1\%. 
We take into account in our analysis this major source of systematic error,
as given by the experiments.

The forward-backward asymmetry can be affected by the interference
between initial-state and final-state radiation, which is known
to change the form of the differential cross section. This effect
is estimated using the program {\tt TOPAZ0}. If we switch the
interference on, $A_{FB}$ is {\em increasing} by 0.1\% for 
an acollinearity cut of 10$^{\circ}$ and by 0.07\% for
an acollinearity cut of 25$^{\circ}$. So the effects are very
small, and as the experiments use differential efficiency
in the scattering angle, the impact on the measured asymmetry values
is much smaller. We can conclude that the interference is under
control.

Another interesting point are the preliminary results on contact
interactions presented by DELPHI~\cite{de209}, L3~\cite{l3209newph}
and OPAL~\cite{op209}. The deviations of $\varepsilon$ from the
SM value are {\em positive} in all cases and amount to 0.8,
0.7 and 1.4 standard deviations respectively. All three
collaborations observe a pull in the same direction, but their
sensitivity is not enough to make any conclusion.
In~\cite{Bourilkov:2000} a limit of $\Lambda_+ > 10.4$~TeV at
95\% confidence level is obtained, which covers about
40\% of the favored region from this analysis at higher energies.

\begin{figure}
  \begin{center}
\mbox {\epsfig{file=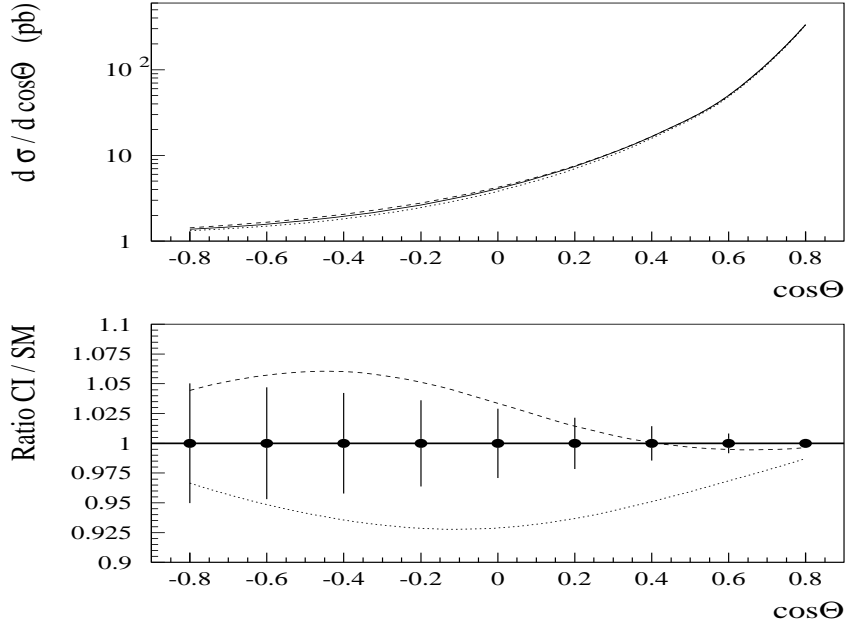,width=0.70\textwidth,height=0.40\textheight}}
  \end{center}
\caption{Effects of contact interactions on the differential cross section
         for Bhabha scattering for centre-of-mass energy 206.8 GeV 
         (upper plot) and the ratio to the Standard Model predictions (lower
         plot): SM - solid, VV model - dotted and
         AA model - dashed. The energy scale is $\Lambda_+ = 10.3$~TeV in
         both cases. The ``data'' points show the expected statistical error
         if we combine the measurements of the four LEP collaborations
         for the six energy points from 192 to 208 GeV.}
\label{figure3}
\end{figure}

In order to clarify the situation, the final results of the four
LEP collaborations are needed. The best option is to combine the
measurements of the differential cross sections in the four
experiments, as illustrated in Fig.~\ref{figure3}. The AA model
gives a specific signature, enhancing the cross section in the
central part and backwards, and only slightly reducing the forward
peak.

When interpreting the physical meaning of contact interaction scales,
we should remember that a strong coupling $\frac{g^2}{4\pi}=1$
for the novel interactions is postulated by convention.
If we assume a coupling of different strength, $\frac{g^2}{4\pi}=\alpha'$,
the limits can be translated as
\mbox{$\Lambda' = \sqrt{\alpha'}\cdot \Lambda$}.
For the extreme case of a coupling of electromagnetic strength
the real scale can be around 1 TeV.

The measurements of Bhabha scattering above the Z resonance
reach already a higher level of precision than the best
theoretical tools available.
In order to fully exploit the physics potential of the large data samples
collected during the LEP running from 1997 to 2000,
improved theory predictions are very desirable.
A combined effort of the four LEP collaborations is needed to answer the
question: is Bhabha scattering opening the first window
to new physics at the TeV scale?

%

\bibliographystyle{myprsty}
\bibliography{%
dopir01}

%
%


%
%

\mediumtext
\begin{table}[h] 
 \renewcommand{\arraystretch}{1.2}
   \caption{
    Results of contact interaction fits to Bhabha scattering.
    The numbers in brackets are the values of 
    $[\eta_{LL}, \eta_{RR},\eta_{LR},\eta_{RL}]$
    defining to which helicity amplitudes the contact interaction contributes.
    The models cover the interference
    of contact terms with single  as well as
    with a combination of helicity amplitudes.  
    The one-sided 95\% confidence level lower limits on the parameters
    $\Lambda_{+}$ ($\Lambda_{-}$) given in TeV  
    correspond to the upper (lower) sign of 
    the parameters $\eta$, respectively.  
    }
  \label{tab:ci-leptons}
    \begin{tabular}{ccrcc}
~~Model~~  & Amplitudes & \multicolumn{1}{c}{$\varepsilon$} &~~~~$\Lambda_-$~~~&~~~~$\Lambda_+$~~~ \\ 
           &$[\eta_{LL}, \eta_{RR},\eta_{LR},\eta_{RL}]$
                        & \multicolumn{1}{c}{[TeV$^{-2}$]}  &~~~~[TeV]~~      &~~~~[TeV]~~~ \\
\hline
~~~~LL~~~~ & $\rm[\pm 1,     0,     0,     0]$ &  0.0028$\;^{+ 0.0066}_{- 0.0055}$ & 10.3 &  8.3  \\
RR         & $\rm[    0, \pm 1,     0,     0]$ &  0.0036$\;^{+ 0.0063}_{- 0.0065}$ & 10.2 &  8.2  \\
\hline                 
LR         & $\rm[    0,     0, \pm 1,     0]$ & -0.0046$\;^{+ 0.0044}_{- 0.0046}$ &  8.8 & 12.7  \\
RL         & $\rm[    0,     0,     0, \pm 1]$ & -0.0046$\;^{+ 0.0044}_{- 0.0046}$ &  8.8 & 12.7  \\
\hline                 
VV         & $\rm[\pm 1, \pm 1, \pm 1, \pm 1]$ & -0.0008$\;^{+ 0.0017}_{- 0.0011}$ & 18.0 & 21.3  \\
AA         & $\rm[\pm 1, \pm 1, \mp 1, \mp 1]$ &  0.0095$\;^{+ 0.0036}_{- 0.0037}$ & 16.5 &  8.0  \\
\hline                 
LL$+$RR    & $\rm[\pm 1, \pm 1,     0,     0]$ &  0.0020$\;^{+ 0.0029}_{- 0.0035}$ & 14.3 & 11.9  \\
LR$+$RL    & $\rm[    0,     0, \pm 1, \pm 1]$ & -0.0023$\;^{+ 0.0023}_{- 0.0024}$ & 12.4 & 18.2  \\
\end{tabular}
\end{table}

\end{document}